\documentclass[aps,prd,twocolumn,nofootinbib]{revtex4-1}

\usepackage{epsfig}
\usepackage{graphicx}
\usepackage{slashbox}

\begin{document}

\title{Hadronic Production of the Doubly Heavy Baryon $\Xi_{bc}$ at LHC}

\author{Jia-Wei Zhang}
\author{Xing-Gang Wu}
\email{wuxg@cqu.edu.cn}
\author{Tao Zhong}
\author{Yao Yu}
\author{Zhen-Yun Fang}

\affiliation{Department of Physics, Chongqing University, Chongqing 400044, People's Republic of China}

\begin{abstract}

We investigate the hadronic production of the doubly heavy baryon $\Xi_{bc}$ at the large hadron collider (LHC), where contributions from the four $(bc)$-diquark states $(bc)_{\bf\bar{3},6}[^1S_0]$ and $(bc)_{\bf\bar{3},6}[^3S_1]$ have been taken into consideration. Numerical results show that under the condition of $p_T>4$ GeV and $|y|<1.5$, sizable $\Xi_{bc}$ events about $ 1.7\times 10^7$ and $3.5\times10^9$ per year can be produced for the center-of-mass energy $\sqrt{S}=7$ TeV and $\sqrt{S}=14$ TeV respectively. For experimental usage, the total and the interested differential cross-sections are estimated under some typical $p_T$- and $y$- cuts for the LHC detectors CMS, ATLAS and LHCb. Main uncertainties are discussed and a comparative study on the hadronic production of $\Xi_{cc}$, $\Xi_{bc}$ and $\Xi_{bb}$ at LHC are also presented. \\



\end{abstract}

\maketitle

\section{Introduction}

The doubly heavy baryons, which represent a new type of objects in comparison with the ordinary baryons, were first predicted by Ref.\cite{quark}. These baryons shall offer a good platform for testing various theories and models, such as the quark model, the perturbative Quantum Chromodynamics (pQCD), the nonrelativistic QCD (NRQCD), the potential model and so on. Moreover, to know these baryons well can help us to understand the heavy-flavor physics, the weak interaction, the charge-parity violation, and etc.. A number of heavy baryons were discovered by several experiment collaborations, such as CLEO, Belle and BaBar, ARGUS, SELEX and CDF collaborations, a review on this point can be found in Refs.\cite{bspect,lhc1,lhc2}. However, for the family of double-heavy baryons, only $\Xi_{cc}$ has been observed and reported \cite{exp1,exp2,exp3,exp4}. While due to their smaller production rate, few $\Xi_{bc}$ and $\Xi_{bb}$ have been observed. Even for $\Xi_{cc}$, its measured production rate and decay width are much larger than most of the theoretical predictions \cite{xicc1,xicc2,xicc3,xicc4,xicc5,xicc6,xicc7,fpro1,fpro2,fpro3}. More data are needed to clarify the present situation.

The CERN Large Hadron Collider (LHC), which is designed to run with a high center-of-mass (C.M.) collision energy up to $14$ TeV and a high luminosity up to $10^{-34}cm^{-2}s^{-1}$ \cite{LHC}, shall be of great help for the purpose. At the present, it is setting up for running with $\sqrt{S}=7$ TeV and will result in an integrated luminosity of 10 $fb^{-1}$ after its first year of running. Taking into account the prospects of observation and measurement of doubly heavy baryons at LHC, it would be interesting to investigate the properties of these states. In the present paper, we shall first concentrate our attention on the hadronic production of $\Xi_{bc}$, and then make a comparative study with those of $\Xi_{cc}$ and $\Xi_{bb}$.

\begin{figure*}
\centering
\includegraphics[width=0.4\textwidth]{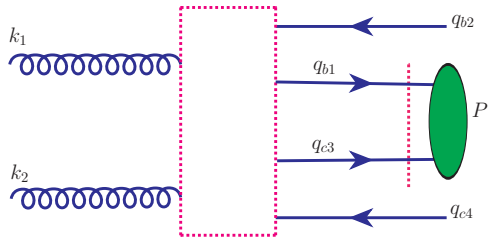}
\caption{Schematic diagram for the hadroproduction of $\Xi_{bc}$ from the gluon-gluon fusion mechanism $g(k_1)+g(k_2) \to\Xi_{bc}(P)+\bar{b}(q_{b2})+\bar{c}(q_{c2})$, where the dashed box stands for the hard interaction kernel.} \label{diag}
\end{figure*}

The doubly heavy baryon can be regarded as a combination of the heavy diquark and a light quark \cite{twobody}. The dominant mechanism for the hadronic production of $\Xi_{bc}$ baryon is the gluon-gluon fusion mechanism via the process $g+g\to\Xi_{bc}+\bar{b}+\bar{c}$. The gluon-gluon fusion mechanism includes 36 Feynman diagrams similar to the case of the $\Xi_{cc}$ baryon production \cite{xicc1,xicc2,xicc3,xicc4,xicc5,xicc6,xicc7,genxicc1,genxicc2} and the $B_c$ meson production \cite{bc1,bc2,bc3}, all of which can be schematically represented by Fig.(\ref{diag}), where $k_1$ and $k_2$ are two momenta for the initial gluons, $q_{b2}$ and $q_{c4}$ are momenta for the two outgoing $\bar{b}$ and $\bar{c}$, $P$ is the momentum of $\Xi_{bc}$. The intermediate $(bc)$-diquark pair can be in one of the four Fock states, i.e. $(bc)_{\bf\bar{3}}[^3S_1]$, $(bc)_{\bf 6}[^1S_0]$, $(bc)_{\bf 6}[^3S_1]$ and $(bc)_{\bf\bar{3}}[^1S_0]$ respectively. More definitely, according to Refs.\cite{xicc6,genxicc1,genxicc2}, the hadronic production of $\Xi_{bc}$ can be divided into three steps: the first step is the production of a $c\bar{c}$-pair and a $b\bar{b}$-pair that can be calculated by pQCD, since the intermediate gluon should be hard enough to form a heavy quark-antiquark pair. The second step is that the two heavy quarks fusion into a binding $(bc)$-diquark, and the third step is the fragmentation of such diquark into the desired baryon by grabbing a light quark and suitable number of gluons when needs. The second and the third steps are non-perturbative, which can be described by a universal matrix element within the NRQCD framework \cite{nrqcd}.

The paper is organized as follows. In Sec.II, we give the main idea in dealing with the $\Xi_{bc}$ hadroproduction. Numerical results are presented in Sec.III. And in Sec.IV, we make a discussion on the main uncertainties for $\Xi_{bc}$ hadroproduction and a comparison of the hadronic production of $\Xi_{bc}$, $\Xi_{cc}$ and $\Xi_{bb}$. The final section is reserved for a summary.

\section{Calculation technology}

Within the NRQCD framework, the total hadronic cross-section for the gluon-gluon fusion mechanism can be schematically written as the following factorization form
\begin{displaymath}
\sigma=F^{g}_{H_{1}}(x_{1},\mu_F) F^{g}_{H_{2}}(x_{2},\mu_F) \bigotimes \hat{\sigma}_{gg \rightarrow \Xi_{bc}}(x_{1},x_{2},\mu_F,\mu_R),
\end{displaymath}
where $F^{i}_{H}(x,\mu_F)$ (with $H=H_1$ or $H_2$; $x=x_1$ or $x_2$) is the distribution function of parton $i$ in hadron $H$. $\mu_F$ is the factorization scale and $\mu_R$ is the renormalization scale, and for convenience, we take them to be the transverse mass of $\Xi_{bc}$, i.e. $\mu_R=\mu_F=\sqrt{M^2_{\Xi_{bc}}+p_T^2}$. $\hat\sigma_{gg\rightarrow \Xi_{bc}}$ stands for the cross-section for the gluon-gluon fusion subprocess, which can be expressed as \cite{xicc7,genxicc1,fpro3},
\begin{widetext}
\begin{eqnarray}
\hat{\sigma}_{gg\rightarrow \Xi_{bc}} &=& H(gg\to (bc)_{\bf\bar 3}[^3S_1] ) \cdot h^{(bc)}_3 +H(gg\to (bc)_{\bf 6}[^1S_0] ) \cdot h^{(bc)}_1 + H(gg\to (bc)_{\bf\bar 3}[^1S_0] ) \cdot {h'}^{(bc)}_3 \nonumber\\
&& +H(gg\to (bc)_{\bf 6}[^3S_1])\cdot {h'}^{(bc)}_1 +\cdots ,
\end{eqnarray}
\end{widetext}
where the ellipsis stands for the terms in higher orders of $v$, $v$ is the relative velocity between the constitute $b$ and $c$ quarks. $H(gg\to (bc)_{\bf\bar{3},6}[^3S_1] )$ or $H(gg\to (bc)_{\bf\bar{3},6}[^1S_0])$ is the perturbative coefficient for producing $(bc)$-diquark in different spin and color configurations respectively. Four matrix elements: $h^{(bc)}_1$, $h^{(bc)}_3$, ${h'}^{(bc)}_1$ and ${h'}^{(bc)}_3$ characterize the transitions of the $(bc)$-diquark in $[^1S_0]_{\bf 6}$, $[^3S_1]_{\bf\bar 3}$, $[^3S_1]_{\bf 6}$, $[^1S_0]_{\bf\bar 3}$ spin and color configurations into $\Xi_{bc}$ baryon respectively. $h^{(bc)}_3$ can be related to the wavefunction of the color anti-triplet diquark $(bc)_{\bf\bar{3}}[^3S_1]$ as $h^{(bc)}_3=|\Psi_{bc}(0)|^2$. According to the discussions shown by Ref.\cite{fpro3}, other matrix elements $h^{(bc)}_1$, ${h'}^{(bc)}_1$ and ${h'}^{(bc)}_3$ are of the same order in $v$ as $h^{(bc)}_3$. Since all these matrix elements emerge as overall parameters, we can easily improve our numerical results when we know these matrix elements well. Naively, we take all of them to be $h^{(bc)}_3$ to do our estimation \cite{genxicc1,genxicc2,fpro3}.

To derive analytical squared amplitude of the 36 Feynman diagrams for the hard subprocess is a tedious task, since it contains non-Abelian gluons and massive fermions. In Refs.\cite{bc2,xicc6}, the so-called improved helicity amplitude approach has been adopted to derive analytic expressions for the process at the amplitude level. And basing on the obtained sententious and analytical expressions, an effective generator GENXICC \cite{genxicc1,genxicc2} for simulating $\Xi_{cc}$, $\Xi_{bc}$ and $\Xi_{bb}$  events has been accomplished. Here we shall use GENXICC to make a detailed study on the hadronic production of $\Xi_{bc}$.

\section{Numerical results}

To be consistent with the leading-order (LO) hard scattering amplitude, the LO parton distribution function (PDF) of CTEQ group, i.e. CTEQ6L \cite{6lcteq}, and the LO running $\alpha_s$ are adopted in doing the numerical calculation. And for other parameters we adopt the following values \cite{xicc2}:
\begin{eqnarray}\label{para1}
&& m_c=1.8GeV,\; m_b=5.1GeV,\nonumber\\
&& M_{\Xi_{bc}}=6.9GeV,\; |\Psi_{bc}(0)|^2=0.065GeV^3 \ .
\end{eqnarray}

\begin{table*}
\caption{Hadronic cross section (in unit $nb$) for $\Xi_{bc}$ at LHC with $\sqrt{S}=7.0$ TeV. Three typical $p_T$ cuts are adopted. As for the rapidity and pseudo-rapidity cut, we take $|y|< 1.5$ and $|y|< 2.5$ for CMS and ATLAS, and $1.9<|\eta|< 4.9$ for LHCb. } \vspace{2mm}
\begin{tabular}{|c|c|c|cc|c|}
\hline\hline
-&-&-& \multicolumn{2}{c|}{~~LHC (CMS, ATLAS)~~}& LHCb \\
\hline -&\backslashbox{$p_{Tcut}$} {$y_{cut}$ or $\eta_{cut}$}& NO cut &$|y|< 1.5$ & $|y|< 2.5$&  $1.9\leq|\eta|\leq 4.9$ \\
\hline $(bc)_{\bf\bar 3}[^3S_1]$&$0$ GeV & 20.90 &10.82 & 16.21&11.12\\
- &$2.5$ GeV & 16.01 &8.363 & 12.49&7.941\\
- &$4.0$ GeV & 10.72 &5.674 & 8.446&4.886\\
\hline $(bc)_{6}[^1S_0]$&$0$ GeV  &5.120 &2.618&3.938&2.689\\
  -&$2.5$ GeV & 4.062 &2.094 &3.142&2.006\\
  -&$4.0$ GeV& 2.853 &1.489 & 2.227&1.307\\
\hline  $(bc)_{6}[^3S_1]$ &$0$ GeV &31.70 &16.03 & 24.20&17.09\\
 - &$2.5$ GeV& 24.09 &12.30 & 18.52&12.17\\
 - &$4.0$ GeV & 15.97 &8.276 & 12.42&7.441\\
\hline  $(bc)_{\bf\bar 3}[^1S_0]$&$0$ GeV  & 5.502&2.886 & 4.315&2.896\\
 - &$2.5$ GeV& 4.280 &2.259 & 3.372&2.096\\
 - &$4.0$ GeV& 2.941 &1.569 & 2.337&1.325\\
\hline\hline
\end{tabular}\label{cross7tev}
\end{table*}

\begin{table*}
\caption{Hadronic cross section (in unit $nb$) for $\Xi_{bc}$ at LHC with $\sqrt{S}=14.0$ TeV. Three typical $p_T$ cuts are adopted. As for the rapidity and pseudo-rapidity cut, we take $|y|< 1.5$ and $|y|< 2.5$ for CMS and ATLAS, and $1.9<|\eta|< 4.9$ for LHCb.} \vspace{2mm}
\begin{tabular}{|c|c|c|cc|c|}
\hline\hline
-&-&-& \multicolumn{2}{c|}{~~LHC(CMS, ATLAS)~~}& LHCb \\
\hline -&\backslashbox{$p_{Tcut}$} {$y_{cut}$ or $\eta_{cut}$}& NO cut &$|y|< 1.5$ &$|y|< 2.5$&  $1.9<|\eta|< 4.9$ \\
\hline $(bc)_{\bf\bar 3}[^3S_1]$ &$0$ GeV& 47.24 &21.70 & 33.43&25.85\\
  -&$2.5$ GeV & 36.55 &16.92 & 26.04 &19.17\\
  - &$4.0$GeV & 24.92 &11.70 & 17.95&12.34\\
\hline  $(bc)_{\bf 6}[^1S_0]$ &$0$ GeV& 11.55 &5.259 &8.112&6.250\\
 -&$2.5$ GeV & 9.255 &4.243 &6.537&4.822\\
 - &$4.0$ GeV& 6.607 &3.067 & 4.713&3.269\\
\hline $(bc)_{\bf 6}[^3S_1]$&$0$ GeV  &70.67 &31.80 & 49.19&38.89\\
  -&$2.5$ GeV& 54.29 &24.65 & 38.07&28.74\\
 -&$4.0$ GeV & 36.59 &16.85 & 25.97&18.36\\
\hline $(bc)_{\bf\bar 3}[^1S_0]$&$0$ GeV  & 12.46 &5.794 & 8.909&6.788\\
  -&$2.5$ GeV& 9.802 &4.591 & 7.049&5.111\\
  -&$4.0$ GeV& 6.855 &3.248 &4.975&3.377\\
\hline\hline
\end{tabular}\label{cross14tev}
\end{table*}

In TAB.\ref{cross7tev} and TAB.\ref{cross14tev}, we show the total cross sections for $\Xi_{bc}$ with its $(bc)$-diquark in $(bc)_{\bf\bar{3},\bf{6}}[^1S_0]$ and $(bc)_{\bf\bar{3},\bf{6}}[^3S_1]$ states respectively. In these two tables, the results for the C.M. energies $\sqrt{S}=7.0$ TeV and $\sqrt{S}=14.0$ TeV are presented. Total cross sections with typical cuts for ATLAS, CMS and LHCb are adopted \cite{ATLAS,CMS,LHCb}, e.g. the transverse momentum cut $p_{tcut}=0$, $2.5$ GeV and $4.0$ GeV, and the rapidity cut $|y|< 1.5$ and $|y|< 2.5$ for ATLAS and CMS, and $1.9< |\eta|< 4.9$ for LHCb are used for the estimation. For CMS, it usually adopts the pseudo-rapidity cut condition around $|\eta|<2.5$, since the $p_T$- and $y$- differential distributions for the four diquark states $(bc)_n$ under the two cases of $|y|< 2.5$ and $|\eta|< 2.5$ are close in shape, and their corresponding total cross sections with pseudo-rapidity cut $|\eta|< 2.5$ are also close to those with rapidity cut $|y|< 2.5$, i.e. $\sigma^{n}_{|\eta|<2.5}/\sigma^{n}_{|y|<2.5} \sim 70\%- 85\%$, so to short the paper we take the same cut conditions for both ATLAS and CMS. Here the short notation $(bc)_{n}$ with (n=1,2,3,4) stands for $(bc)_{\bf\bar{3}}[^1S_0]$, $(bc)_{\bf{6}}[^1S_0]$, $(bc)_{\bf\bar{3}}[^3S_1]$ and $(bc)_{\bf{6}}[^3S_1]$ respectively.

\begin{figure*}
\centering
\includegraphics[width=0.4\textwidth]{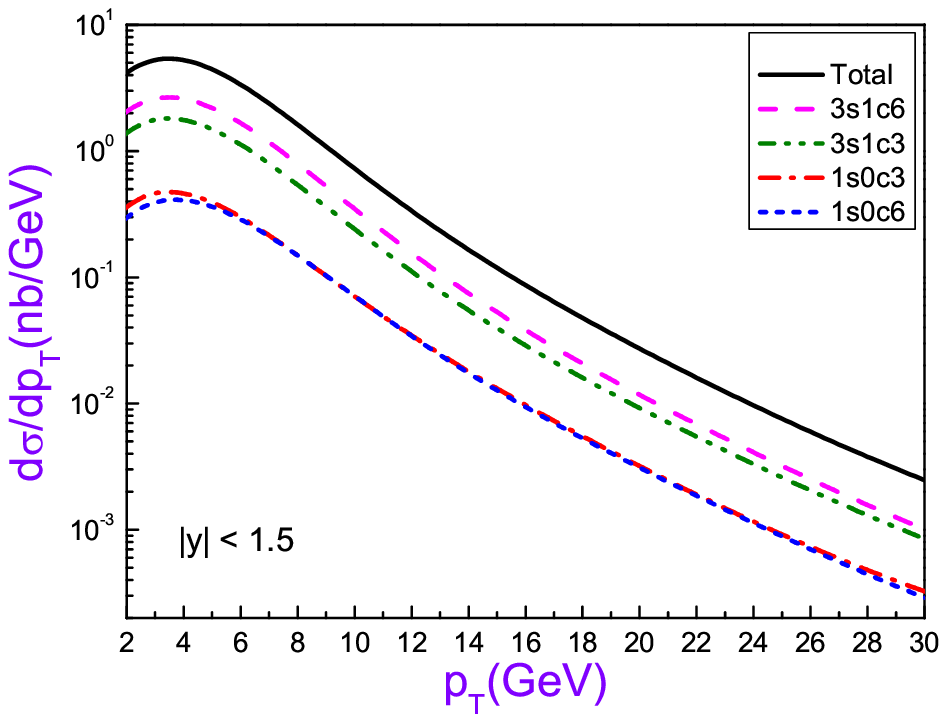}
\includegraphics[width=0.4\textwidth]{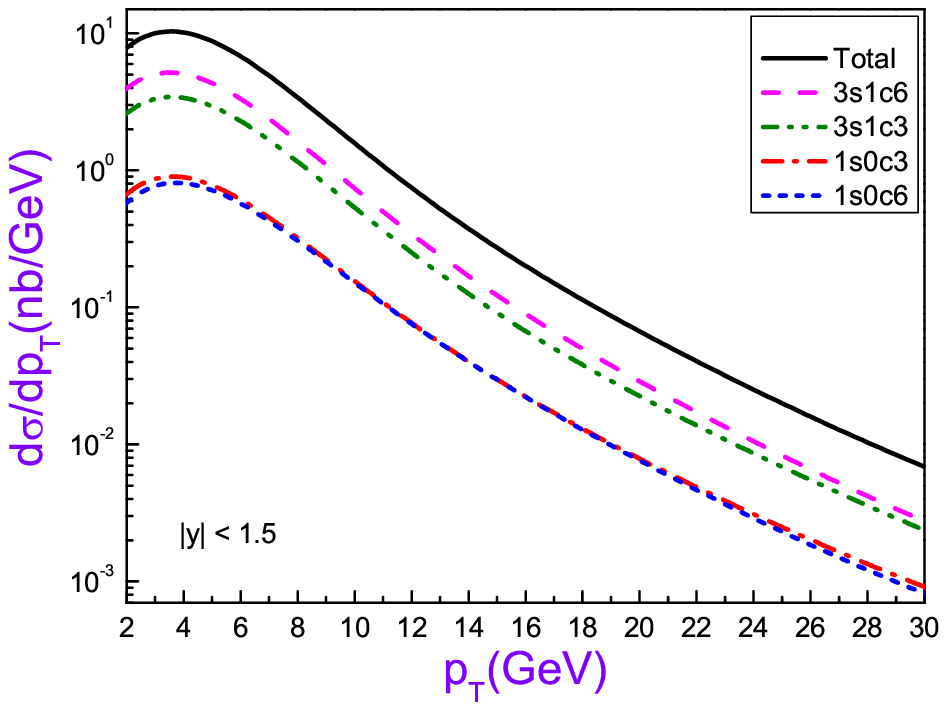}
\caption{$p_{T}$-distributions for $\Xi_{bc}$ hadroproduction under ATLAS and CMS rapidity cut $|y|< 1.5$, where the left and the right diagrams are for $\sqrt{S}=7$ TeV and  $\sqrt{S}=14$ TeV respectively. The solid, the dashed, the dash-dot-dot, the dash-dot and the short-dash lines stand for the total, that of $(bc)_{\bf 6}[^3S_1]$, $(bc)_{\bf\bar{3}}[^3S_1]$, $(bc)_{\bf\bar{3}}[^1S_0]$ and $(bc)_{\bf 6}[^1S_0]$ respectively.} \label{y15}
\end{figure*}

\begin{figure*}
\centering
\includegraphics[width=0.4\textwidth]{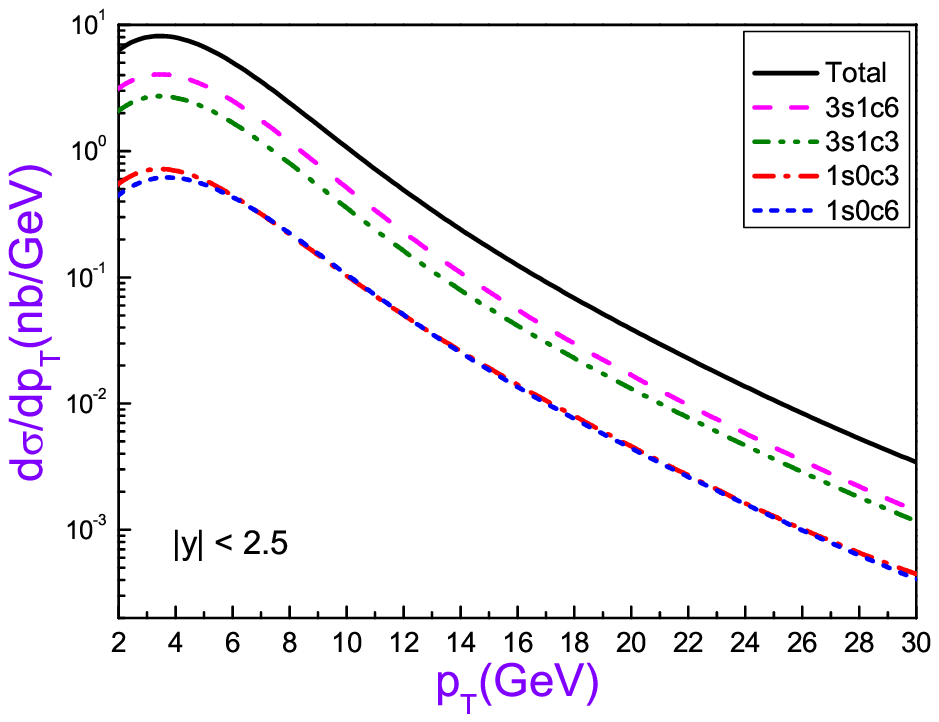}
\includegraphics[width=0.4\textwidth]{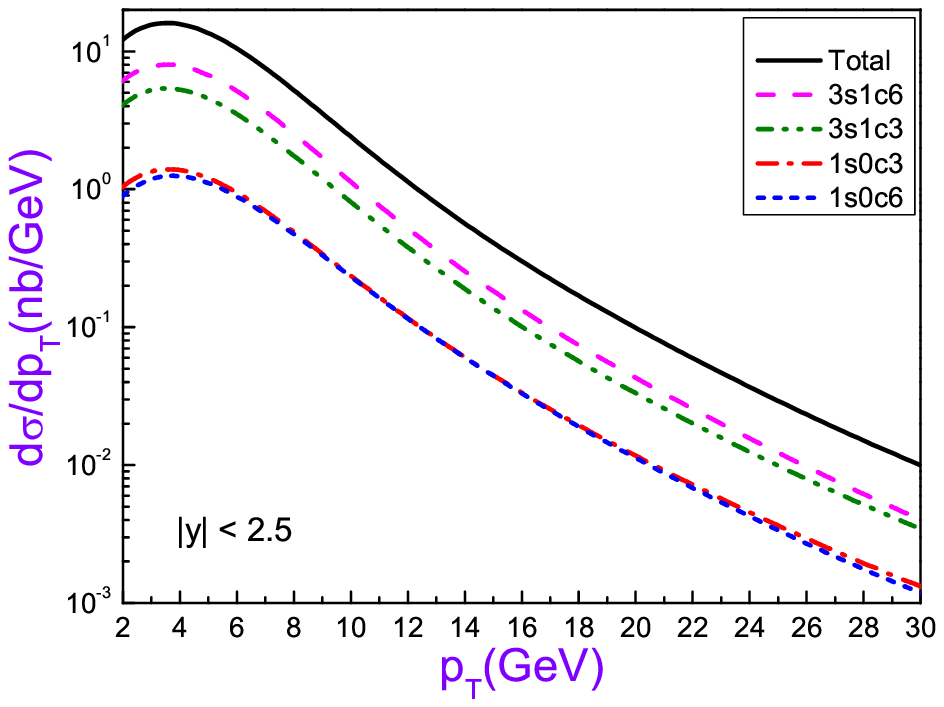}
\caption{$p_{T}$-distributions for $\Xi_{bc}$ hadroproduction under ATLAS and CMS rapidity cut $|y|< 2.5$, where the left and the right diagrams are for $\sqrt{S}=7$ TeV and  $\sqrt{S}=14$ TeV respectively. The solid, the dashed, the dash-dot-dot, the dash-dot and the short-dash lines stand for the total, that of $(bc)_{\bf 6}[^3S_1]$, $(bc)_{\bf\bar{3}}[^3S_1]$, $(bc)_{\bf\bar{3}}[^1S_0]$ and $(bc)_{\bf 6}[^1S_0]$ respectively.} \label{y25}
\end{figure*}

\begin{figure*}
\centering
\includegraphics[width=0.4\textwidth]{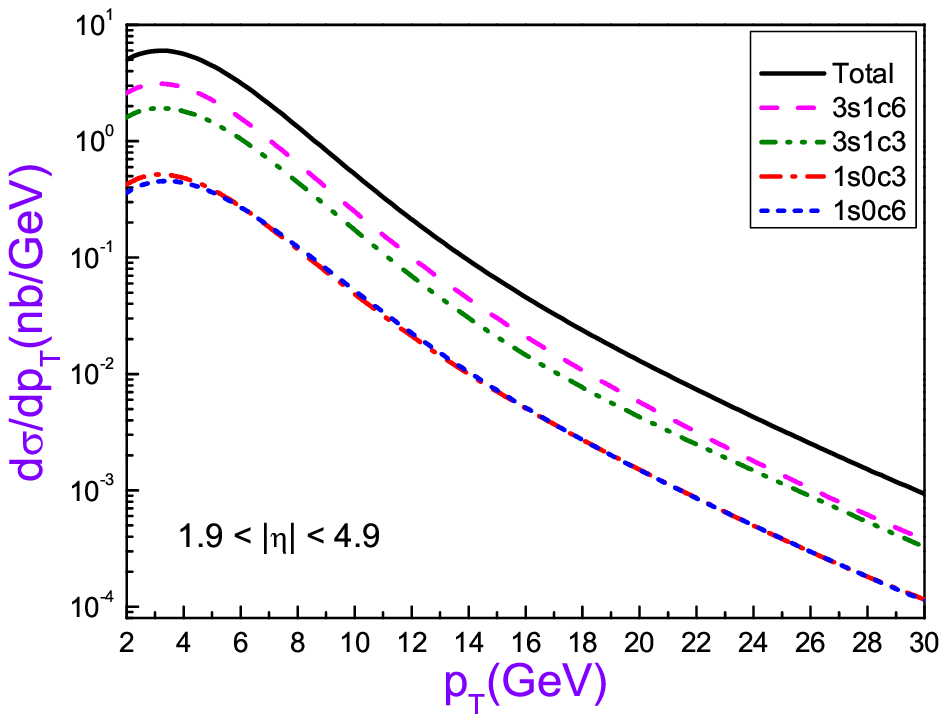}
\includegraphics[width=0.4\textwidth]{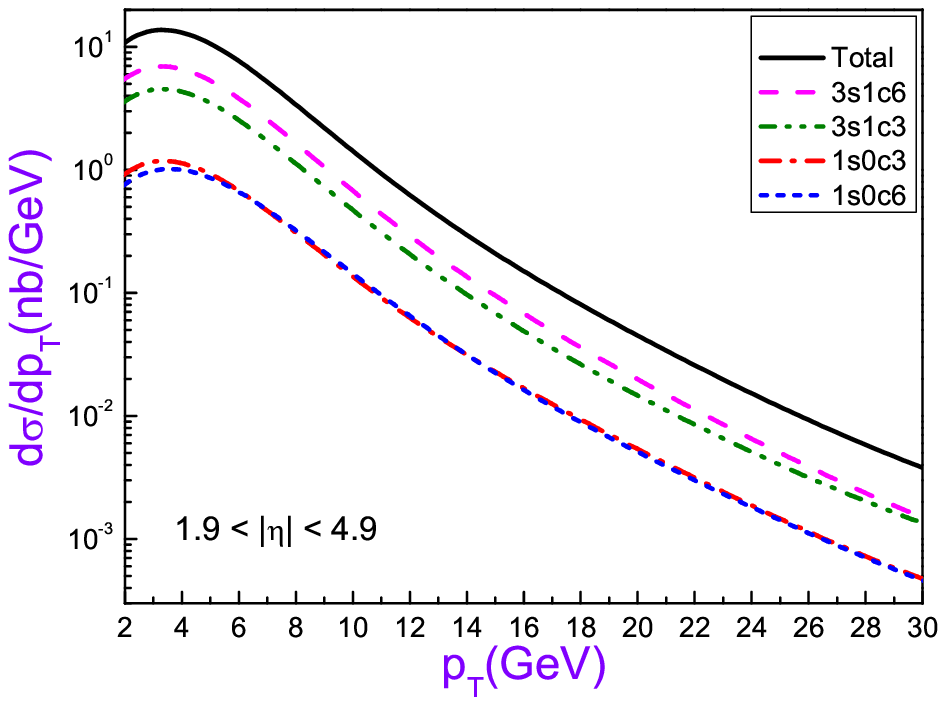}
\caption{$p_{T}$-distributions for $\Xi_{bc}$ hadroproduction under LHCb pseudo-rapidity cut $1.9<|\eta|< 4.9$, where the left and the right diagrams are for $\sqrt{S}=7$ TeV and  $\sqrt{S}=14$ TeV respectively. The solid, the dashed, the dash-dot-dot, the dash-dot and the short-dash lines stand for the total, that of $(bc)_{\bf 6}[^3S_1]$, $(bc)_{\bf\bar{3}}[^3S_1]$, $(bc)_{\bf\bar{3}}[^1S_0]$ and $(bc)_{\bf 6}[^1S_0]$ respectively.} \label{eta19}
\end{figure*}

TABs.(\ref{cross7tev},\ref{cross14tev}) show that all the four diquark states $(bc)_{\bf\bar{3},\bf{6}}[^1S_0]$ and $(bc)_{\bf\bar{3},\bf{6}}[^3S_1]$ can provide sizable contributions to $\Xi_{bc}$ hadroproduction. Moreover, one may observe $\sigma_{(bc)_{\bf 6}[^3S_1]}>\sigma_{(bc)_{\bf\bar{3}}[^3S_1]}> \sigma_{(bc)_{\bf\bar{3}}[^1S_0]}\succsim\sigma_{(bc)_{\bf 6}[^1S_0]}$. And the total cross section for the scalar diquark states $(bc)_{\bar{3},6}[^1S_0]$ is about $20\%$ of that of the vector diquark states $(bc)_{\bf\bar{3},\bf{6}}[^3S_1]$. Differential cross sections versus $\Xi_{bc}$-$p_{T}$ are drawn in Figs.(\ref{y15},\ref{y25},\ref{eta19}), where the results for the three typical rapidity or pseudo-rapidity cuts $|y|< 1.5$, $|y|< 2.5$ and $1.9<|\eta|< 4.9$ are presented and these curves show the relative importance of the four diquark states clearly.

The LHC has been first running at $\sqrt{S}=7.0$ TeV with luminosity $2.0 \times 10^{-32}cm^{-2}s^{-1}$ from 30th March 2010, and its integrated luminosity is 10 $fb^{-1}$/yr. Based on the total cross sections shown in TAB.\ref{cross7tev}, one can estimate that about $1.7 \times 10^7$ $\Xi_{bc}$ events per year can be produced under the condition of $p_T>4$ GeV and $|y|<1.5$. When the C.M. energy and the luminosity are reached up to 14 TeV and $10^{-34}cm^{-2}s^{-1}$ as designed, then the integrated luminosity will be changed to 100 $fb^{-1}$/yr, one can estimate that about $3.5\times10^9$ $\Xi_{bc}$ events per year can be produced under the condition of $p_T>4$ GeV and $|y|<1.5$.

\begin{figure*}
\centering
\includegraphics[width=0.4\textwidth]{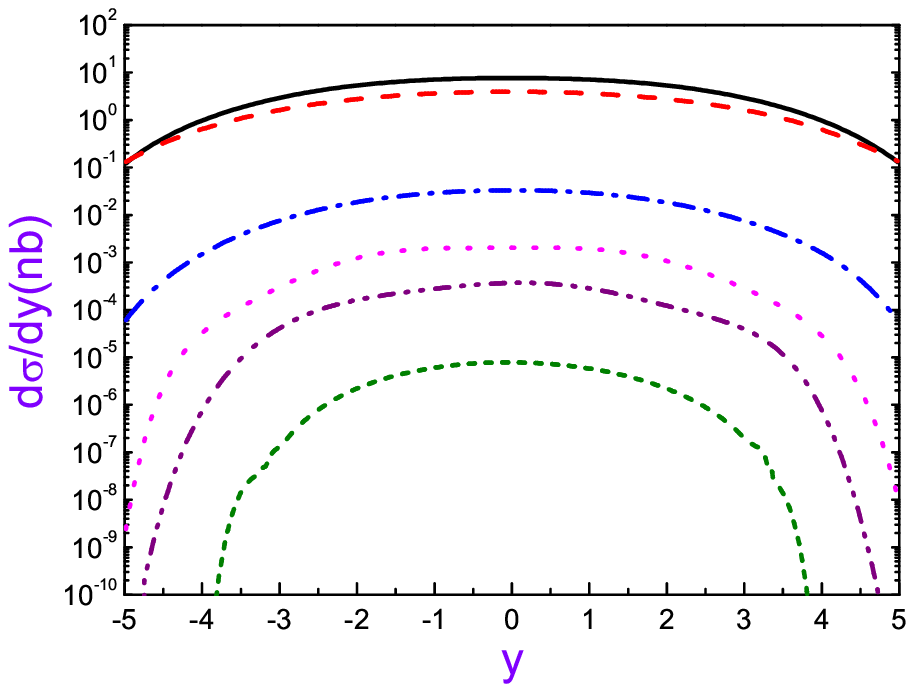}
\includegraphics[width=0.4\textwidth]{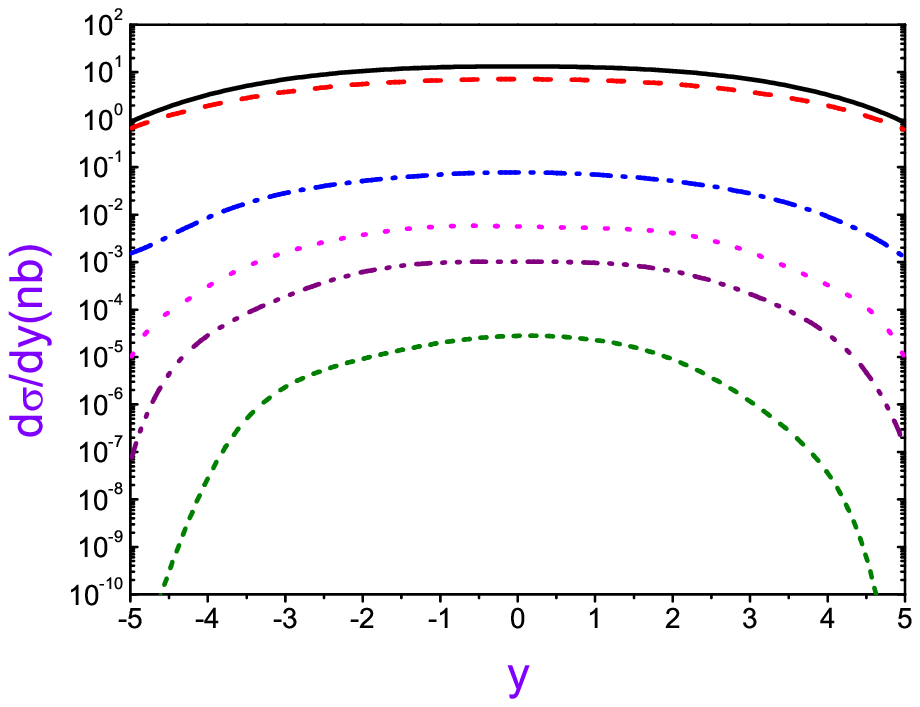}
\caption{The $y$-distributions for $\Xi_{bc}$ production with various $p_{Tcut}$ in LHC , where the left and the right diagrams are for $\sqrt{S}=7.0$ TeV and  $\sqrt{S}=14.0$ TeV. The Solid line corresponds to the full production without $p_{Tcut}$, the dashed, the dash-dot, the dotted, the dash-dot-dot and the short dashed lines are for $p_{Tcut}=4.0$ GeV, $20.0$ GeV, $35.0$ GeV, $50.0$ GeV $100.0$ GeV respectively. All the curves are the sum of all the four diquark states.} \label{ptcut}
\end{figure*}

\begin{figure*}
\centering
\includegraphics[width=0.4\textwidth]{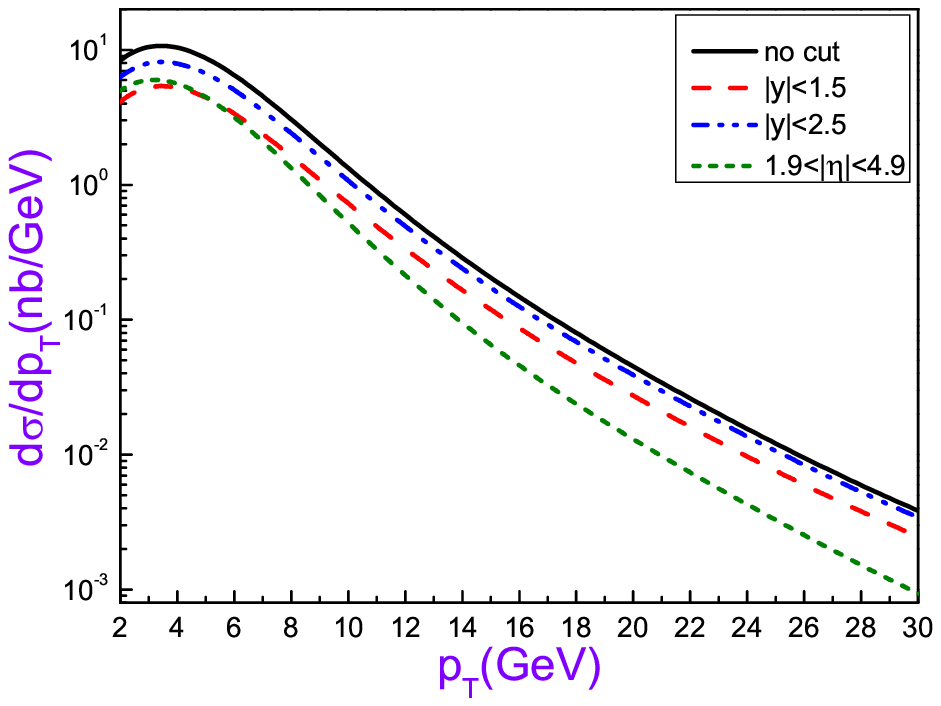}
\includegraphics[width=0.4\textwidth]{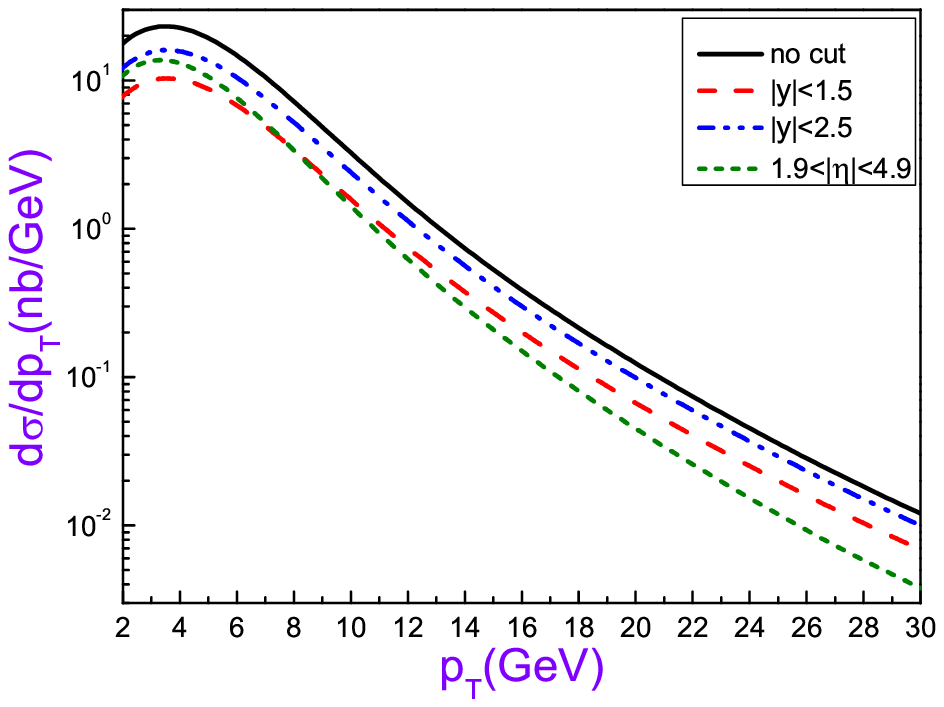}
\caption{The $p_{T}$-distributions for $\Xi_{bc}$ production under various $y$- and $\eta$- cut, where the left and the right diagrams are for $\sqrt{S}=7$ TeV and  $\sqrt{S}=14$ TeV respectively. The solid, the dashed, the dash-dot-dot and the short dash lines stand for no-cut, $|y|< 1.5$, $|y|< 2.5$ and $1.9< |\eta|< 4.9$ respectively. All the curves are the sum of all the four diquark states.} \label{ycut}
\end{figure*}

Next, we draw the $p_{T}$- and $y$- distributions under some typical $p_T$- and $y$- cuts in Figs.(\ref{ptcut},\ref{ycut}), where each curve stands for the sum of all the four diquark states $(bc)_{\bf\bar{3},\bf{6}}[^1S_0]$ and $(bc)_{\bf\bar{3},\bf{6}}[^3S_1]$. The results for $\sqrt{S}=7.0$ TeV and $\sqrt{S}=14.0$ TeV are presented accordingly. Firstly, as shown by Fig.(\ref{ptcut}), there is an obvious platform within the region of $|y|\precsim 3.0$, where dominant contributions to the cross section are there. To show this point clearly, we define a ratio $R^{p_{Tcut}}_c=\left[\sigma_{|y|<c} /\sigma_{tot}\right]_{p_{Tcut}}$, where $\sigma_{tot}$ stands for the total cross section without $y$-cut and $c$ stands for some particular value. Then we obtain $R^{4GeV}_{1.5}=52\%$ and $R^{4GeV}_{2.5}=78\%$ for $\sqrt{S}=7.0$ TeV, and $R^{4GeV}_{1.5}=47\%$ and $R^{4GeV}_{2.5}=72\%$ for $\sqrt{S}=14.0$ TeV. Secondly, as shown by Fig.(\ref{ycut}), the $p_{T}$-distribution under the case of $1.9<|\eta|< 4.9$ drops faster than the cases with other $y$-cuts, especially in the large $p_T$ regions. This implies that if the same larger $p_T$ cut (e.g. $p_t>10$ GeV) is imposed at the colliders \footnote{Since the events move very close to the beam direction cannot be detected by the detectors directly, so such kind of events cannot be utilized for experimental studies in common cases.}, ATLAS and CMS are better than LHCb for studying $\Xi_{bc}$ properties, since more events can be produced and measured at ATLAS and CMS.

\section{Discussions}

\subsection{Main uncertainties for $\Xi_{bc}$ hadroproduction}

To be more useful experimentally, we make a simple discussion on the uncertainties for $\Xi_{bc}$ hadroproduction. For the present LO estimation, the uncertainty sources include the non-perturbative matrix elements, the factorization scale $\mu_F$, the constitute quark masses $m_b$ and $m_c$, PDF and etc..

Numerically, it is found that similar to the hadronic production of $B_c$, $B_s$ and $\Xi_{cc}$ that have been done in the literature, the LO PDFs like MRST2001L \cite{mrst} and CETQ6L \cite{6lcteq} only lead to small difference to the total cross section that is less than $15\%$. So we shall fix the PDF to be CTEQ6L to do our discussion. Moreover, all the non-perturbative matrix elements emerge as overall parameters, then we can easily improve our numerical results when we know these matrix elements well. In the following, we shall concentrate our attention on the uncertainties caused by the factorization scale $\mu_F$, and the constitute quark masses $m_b$ and $m_c$.

\begin{table*}
\caption{Hadronic cross section (in unit $nb$) of $\Xi_{bc}$ at LHC for two typical energy scales A and B. $p_{Tcut}=4.0$ GeV, $|y|< 1.5$ and $|y|< 2.5$ for CMS and ATLAS, and $1.9<|\eta|< 4.9$ for LHCb are adopted for the estimation.} \vspace{2mm}
\begin{tabular}{|c|c|c|c|c|c|c|c|}
\hline
\hline -& $\mu_F$& ~~~A~~~ & ~~~B~~~& ~~~A~~~&~~~B~~~ &~~~A~~~ &~~~B~~~\\
\hline -&\backslashbox{C.M. Energy} {$y_{cut}$ or $\eta_{cut}$}& \multicolumn{2}{c|}{$|y|< 1.5$
} & \multicolumn{2}{c|}{$|y|< 2.5$}& \multicolumn{2}{c|}{$1.9< |\eta|< 4.9$} \\
\hline $(bc)_{\bf\bar 3}[^3S_1]$& $\sqrt{S}=7.0$ TeV  & 3.957 &5.674 & 5.889&8.446 &3.403 &4.886\\
- &$\sqrt{S}=14.0$ TeV  & 8.477&11.70&12.99&17.95 &8.893 &12.34\\
\hline $(bc)_{6}[^1S_0]$& $\sqrt{S}=7.0$ TeV &1.078 &1.489 &1.612  &2.227 &0.939 &1.307\\
-&$\sqrt{S}=14.0$ TeV & 2.298  &3.067 &3.529  &4.713 &2.432 &3.269 \\
\hline  $(bc)_{6}[^3S_1]$ & $\sqrt{S}=7.0$ TeV  & 6.135 & 8.276 & 9.200 &12.42 &5.471 &7.441\\
 - &$\sqrt{S}=14.0$ TeV& 12.90& 16.85 &19.85 &25.97 &13.94 &18.36\\
\hline  $(bc)_{\bf\bar 3}[^1S_0]$& $\sqrt{S}=7.0$ TeV  &1.104& 1.569 &1.646 &2.337 &0.941 & 1.325\\
 - &$\sqrt{S}=14.0$ TeV  &2.360& 3.248 & 3.617& 4.975&2.455 & 3.377\\
\hline\hline
\end{tabular}\label{alphas}
\end{table*}

In TAB.\ref{alphas}, we present the total cross sections for two typical factorization scales, i.e. type A: $\mu_F=\sqrt{\hat{s}/4}$ with $\hat{s}=x_{1} x_{2} S$, and type B: $\mu_F=\sqrt{M^2_{\Xi_{bc}}+p_T^2}$. Other parameters are fixed to be their center values. Here, $p_{Tcut}=4.0$ GeV, and the rapidity cut $|y|< 1.5$ and $|y|< 2.5$ for ATLAS and CMS, and $1.9< |\eta|< 4.9$ for LHCb are adopted for the estimation. It is found that the cross section differences caused by these two factorization scales is about $25\%-30\%$ for the four diquark states $(bc)_{\bf\bar{3},\bf{6}}[^1S_0]$ and $(bc)_{\bf\bar{3},\bf{6}}[^3S_1]$ respectively, which is a comparatively large effect.

Next, we investigate the uncertainties of $m_b$ and $m_c$ in `a factorization way'. More explicitly, when focusing on the uncertainty from $m_b$, we let it be a basic input varying in a possible range $m_b=5.1\pm{0.20}$ GeV with all the other parameters being fixed to their center values, e.g. $m_c=1.80$ GeV, $M_{\Xi_{bc}}=m_b + m_c$ and $\mu_F=\sqrt{M^2_{\Xi_{bc}}+p_T^2}$. Similarly, when discussing the uncertainty caused by $m_c$, we vary $m_c$ within the region of $m_c=1.8\pm{0.10}$ GeV with all the other parameters being fixed to be their center values.

\begin{table*}
\caption{Hadronic cross section (in unit $nb$) of $\Xi_{bc}$ at LHC with varying $m_b \in [4.9,5.3]$ GeV. Other parameters are fixed to be their center values. $p_{Tcut}=4.0$ GeV, $|y|< 1.5$ and $|y|< 2.5$ for CMS and ATLAS, and $1.9<|\eta|< 4.9$ for LHCb are adopted for the estimation.} \vspace{2mm}
\begin{tabular}{|c|c|c|c|c|}
\hline
\hline -&\backslashbox{C.M. Energy} {$y_{cut}$ or $\eta_{cut}$}& ~~~$|y|< 1.5$~~~ & ~~~$|y|< 2.5$~~~ &  ~~~$1.9<|\eta|< 4.9$ ~~~ \\
\hline $(bc)_{\bf\bar 3}[^3S_1]$& $\sqrt{S}=7.0$ TeV  & $5.674^{+0.593}_{-0.519}$ & $8.446^{+0.905}_{-0.769}$&$4.886^{+0.526}_{-0.443}$\\
- &$\sqrt{S}=14.0$ TeV  & $11.70^{+1.15}_{-1.02}$& $17.95^{+1.76}_{-1.58}$ &$12.34^{+1.24}_{-1.10}$ \\
\hline $(bc)_{6}[^1S_0]$& $\sqrt{S}=7.0$ TeV &$1.489^{+0.137}_{-0.122}$ & $2.227^{+0.207}_{-0.185}$&$1.307^{+0.123}_{-0.109}$ \\
-&$\sqrt{S}=14.0$ TeV &  $3.067^{+0.265}_{-0.237}$ & $4.713^{+0.416}_{-0.368}$&$3.269^{+0.292}_{-0.260}$ \\
\hline  $(bc)_{6}[^3S_1]$ & $\sqrt{S}=7.0$ TeV  & $8.276^{+0.868}_{-0.761}$ & $12.42^{+1.31}_{-1.16}$ &$7.441^{+0.797}_{-0.701}$ \\
 - &$\sqrt{S}=14.0$ TeV& $16.85^{+1.67}_{-1.49}$  & $25.97^{+2.58}_{-2.35}$ &$18.36^{+1.83}_{-1.67}$ \\
\hline  $(bc)_{\bf\bar 3}[^1S_0]$& $\sqrt{S}=7.0$ TeV  & $1.569^{+0.167}_{-0.148}$ & $2.337^{+0.250}_{-0.225}$ &$1.325^{+0.144}_{-0.128}$ \\
 - &$\sqrt{S}=14.0$ TeV  & $3.248^{+0.319}_{-0.300}$ & $4.975^{+0.494}_{-0.449}$ & $3.377^{+0.334}_{-0.309}$\\
\hline\hline
\end{tabular}\label{mb}
\end{table*}

\begin{table*}
\caption{Hadronic Cross section (in unit $nb$) of $\Xi_{bc}$ at LHC with varying $m_c \in [1.7,1.9]$ GeV. Other parameters are fixed to be their center values. $p_{Tcut}=4.0$ GeV, $|y|< 1.5$ and $|y|< 2.5$ for CMS and ATLAS, and $1.9<|\eta|< 4.9$ for LHCb are adopted for the estimation.} \vspace{2mm}
\begin{tabular}{|c|c|c|c|c|}
\hline
\hline -&\backslashbox{C.M. Energy} {$y_{cut}$ or $\eta_{cut}$}& ~~~$|y|< 1.5$~~~ & ~~~$|y|< 2.5$~~~ &  ~~~$1.9< |\eta|< 4.9$ ~~~ \\
\hline $(bc)_{\bf\bar 3}[^3S_1]$& $\sqrt{S}=7.0$ TeV  & $5.674^{+1.138}_{-0.893}$ & $8.446^{+1.704}_{-1.330}$&$4.886^{+1.001}_{-0.128}$\\
- &$\sqrt{S}=14.0$ TeV  & $11.70^{+2.31}_{-1.80}$& $17.95^{+3.54}_{-2.79}$ &$12.34^{+2.45}_{-1.96}$ \\
\hline $(bc)_{6}[^1S_0]$& $\sqrt{S}=7.0$ TeV &$1.489^{+0.336}_{-0.260}$ & $2.227^{+0.506}_{-0.389}$&$1.307^{+0.302}_{-0.231}$ \\
-&$\sqrt{S}=14.0$ TeV &  $3.067^{+0.683}_{-0.525}$ & $4.713^{+1.049}_{-0.807}$&$3.269^{+0.733}_{-0.570}$ \\
\hline  $(bc)_{6}[^3S_1]$ & $\sqrt{S}=7.0$ TeV  & $8.276^{+1.611}_{-1.282}$ & $12.42^{+2.41}_{-1.93}$ &$7.441^{+1.468}_{-1.163}$ \\
 - &$\sqrt{S}=14.0$ TeV& $16.85^{+3.20}_{-2.56}$  & $25.97^{+4.95}_{-3.98}$ &$18.36^{+3.53}_{-2.85}$ \\
\hline  $(bc)_{\bf\bar 3}[^1S_0]$& $\sqrt{S}=7.0$ TeV  & $1.569^{+0.307}_{-0.247}$ & $2.337^{+0.458}_{-0.369}$ &$1.325^{+0.278}_{-0.212}$ \\
 - &$\sqrt{S}=14.0$ TeV  & $3.248^{+0.627}_{-0.506}$ & $4.975^{+0.967}_{-0.777}$ & $3.377^{+0.658}_{-0.541}$\\
\hline\hline
\end{tabular}\label{mc}
\end{table*}

We present the total cross sections for $\Xi_{bc}$ with varying $m_b$ or $m_c$ for C.M. energies $\sqrt{S}=7.0$ TeV and $\sqrt{S}=14.0$ TeV in TAB.\ref{mb} and TAB.\ref{mc}. Here, $p_{tcut}=4.0$ GeV, the rapidity cut $|y|< 1.5$ and $|y|< 2.5$ for ATLAS and CMS, and $1.9< |\eta|< 4.9$ for LHCb are adopted for the estimation. Quantitatively, it can be found that the total cross sections decreases with the increment of $m_b$ or $m_c$, which can be roughly explained by the smaller production phase space for larger quark masses. And from TAB.\ref{mb} and TAB.\ref{mc}, one may observe that the cross sections are more sensitive to the value of $m_c$ than $m_b$. When $m_b$ increases or decreases by the step of $0.2$ GeV, the cross section of $\Xi_{bc}$ changes around $8\%-10\%$ for the four diquark states $(bc)_{\bf\bar{3},\bf{6}}[^1S_0]$ and $(bc)_{\bf\bar{3},\bf{6}}[^3S_1]$. While for the case of $m_c$, when $m_c$ increases or decreases by step of $0.1$ GeV, the cross section of $\Xi_{bc}$ decreases or increases by $15\%-20\%$ for the four diquark states $(bc)_{\bf\bar{3},\bf{6}}[^1S_0]$ and $(bc)_{\bf\bar{3},\bf{6}}[^3S_1]$.

\subsection{A comparison of the hadronic production of $\Xi_{cc}$, $\Xi_{bc}$ and $\Xi_{bb}$}

\begin{figure*}
\centering
\includegraphics[width=0.4\textwidth]{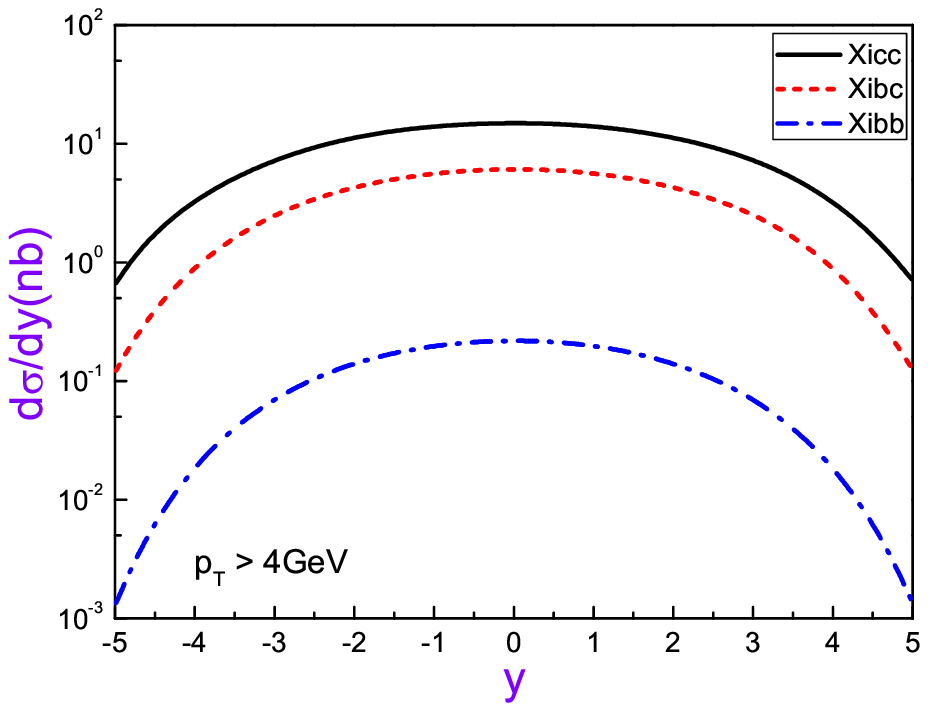}
\includegraphics[width=0.4\textwidth]{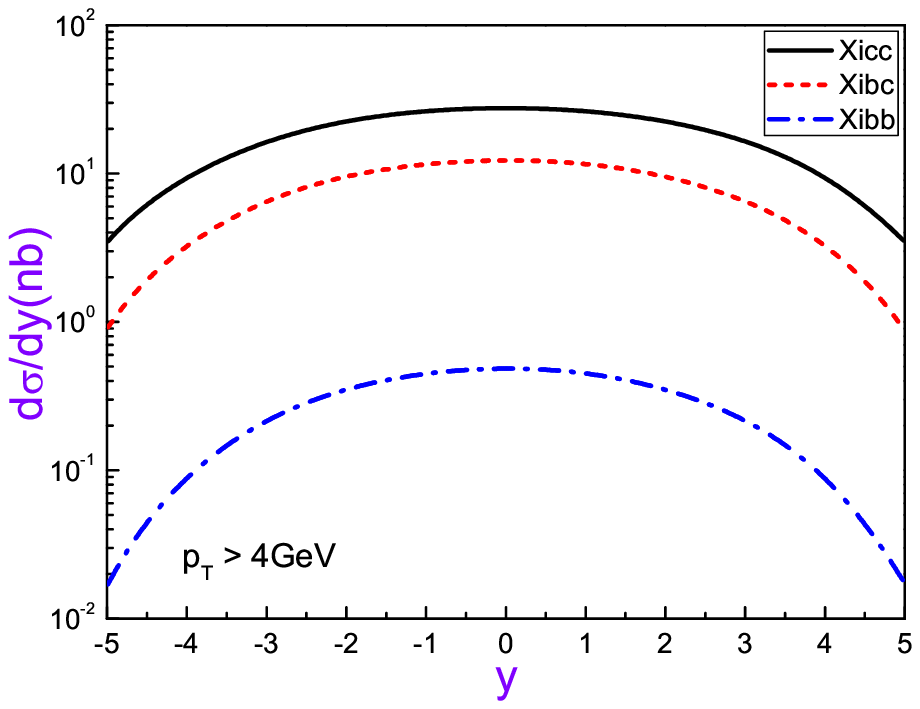}
\caption{The $y$-distributions for $\Xi_{cc}$, $\Xi_{bc}$ and $\Xi_{bb}$ production with $p_T > 4$ GeV in LHC, where the left and the right diagrams are for $\sqrt{S}=7.0$ TeV and  $\sqrt{S}=14.0$ TeV. The Solid, the short dash and the dash-dot lines are for $\Xi_{cc}$, $\Xi_{bc}$ and $\Xi_{bb}$ respectively. All the curves are the sum of all the s-wave diquark states.} \label{pt4yall}
\end{figure*}

\begin{figure*}
\centering
\includegraphics[width=0.4\textwidth]{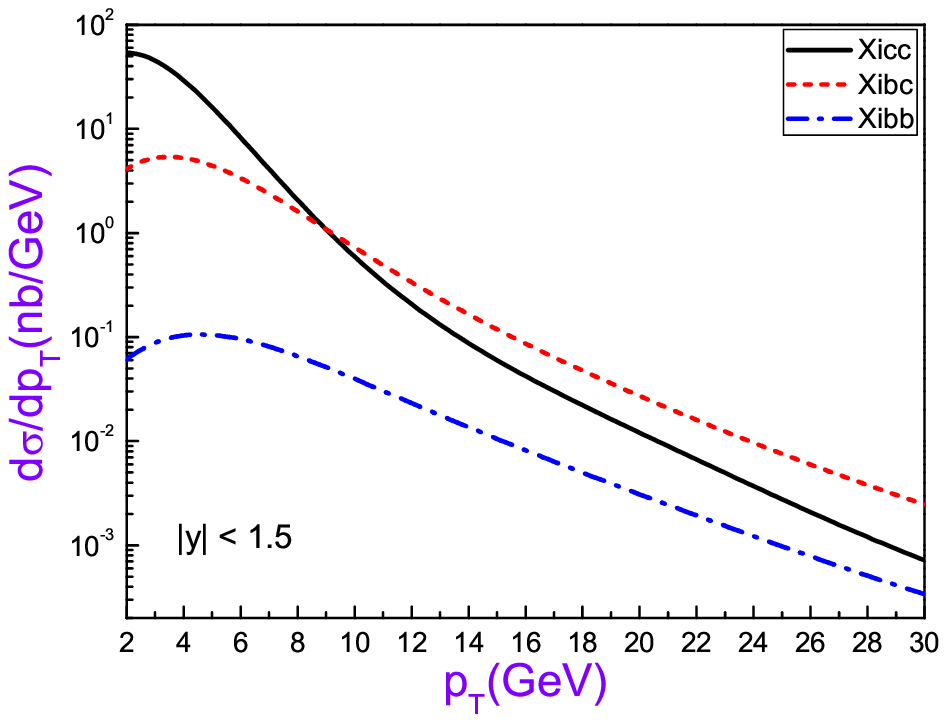}
\includegraphics[width=0.4\textwidth]{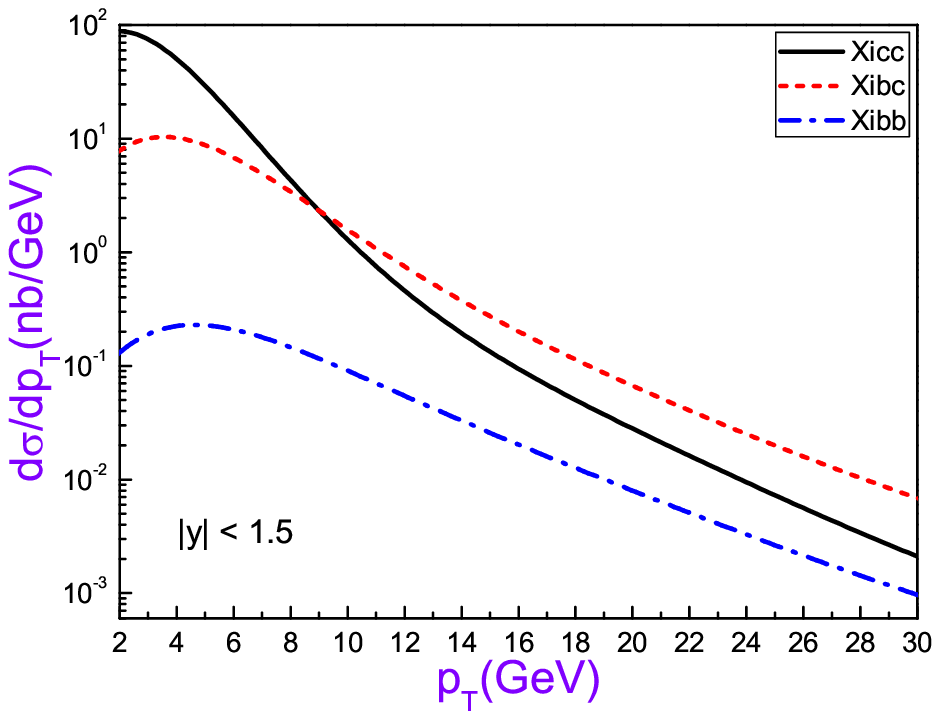}
\caption{The $p_{T}$-distributions for $\Xi_{cc}$, $\Xi_{bc}$ and $\Xi_{bb}$ production with rapidity cut condition $|y|< 1.5$ in LHC, where the left and the right diagrams are for $\sqrt{S}=7.0$ TeV and  $\sqrt{S}=14.0$ TeV. The Solid, the short dash and the dash-dot lines are for $\Xi_{cc}$, $\Xi_{bc}$ and $\Xi_{bb}$ respectively. All the curves are the sum of all the s-wave diquark states.} \label{y15ptall}
\end{figure*}

\begin{table*}
\caption{Comparison of the total cross section (in unit $nb$) for the hadronic production of $\Xi_{cc}$, $\Xi_{bc}$ and $\Xi_{bb}$ at $\sqrt{S}=7.0$ TeV and $\sqrt{S}=14.0$ TeV, where $[^{3}S_1]$ and $[^{1}S_0]$ stand for the combined results for the diquark in spin-triplet and spin-singlet states respectively. In the calculations, we adopt $p_{T}>4$ GeV and $|y|<1.5$.} \vspace{2mm}
\begin{tabular}{|c|c|c|c|c|c|c|}
\hline
-& \multicolumn{2}{c|}{$\Xi_{cc}$} & \multicolumn{2}{c|}{$\Xi_{bc}$}  & \multicolumn{2}{c|}{$\Xi_{bb}$}  \\
\hline
- & $\sqrt{S}=7.0$TeV  & $\sqrt{S}=14.0$TeV  &$\sqrt{S}=7.0$TeV & $\sqrt{S}=14.0$TeV & $\sqrt{S}=7.0$TeV  &$\sqrt{S}=14.0$TeV \\
\hline
$[^{3}S_1]$ & 38.11  & 69.40 &16.7& 28.55  & 0.503&1.137\\
$[^{1}S_0]$ & 9.362 & 17.05 &3.72& 6.315  & 0.100 &0.226\\
\hline
Total & 47.47 & 86.45 &20.42& 34.87  & 0.603 &1.363\\
\hline
\end{tabular}\label{cbccross14tev}
\end{table*}

To be useful reference, we make a comparison of the hadronic production of $\Xi_{cc}$, $\Xi_{bc}$ and $\Xi_{bb}$ at LHC. The total cross sections are presented in TAB.\ref{cbccross14tev}, where $[^{3}S_1]$ and $[^{1}S_0]$ stand for the results for the diquark in spin-triplet and spin-singlet states respectively. More explicitly, for hadronic production of $\Xi_{cc}$, one needs to consider the contributions from the two diquark states $(cc)_{\bf\bf{6}}[^1S_0]$ and $(cc)_{\bf\bar{3}}[^3S_1]$. As for hadronic production of $\Xi_{bb}$, one needs to consider the contributions from the two diquark states $(bb)_{\bf\bf{6}}[^1S_0]$ and $(bb)_{\bf\bar{3}}[^3S_1]$. While for the case of $\Xi_{bc}$, one needs to consider the contributions from the four diquark states $(bc)_{\bf\bar{3},\bf{6}}[^1S_0]$ and $(bc)_{\bf\bar{3},\bf{6}}[^3S_1]$.

From TAB.\ref{cbccross14tev}, one can see that the total cross section of $\Xi_{bc}$ is at the same order of that of $\Xi_{cc}$, i.e. it is about $40\%$ and $44\%$ of that of $\Xi_{cc}$ for $\sqrt{S}=7$ TeV and  $\sqrt{S}=14$ TeV respectively. While, the total cross section of $\Xi_{bb}$ is only $1.5\%$ and $2\%$ of that of $\Xi_{cc}$ for $\sqrt{S}=7$ TeV and $\sqrt{S}=14$ TeV respectively. Then, similar to the case of $\Xi_{cc}$ that has been measured by the SELEX experiment at TEVATRON \cite{exp1,exp2}, it would be possible for $\Xi_{bc}$ be fully studied at LHC.

We draw the $y$- and $p_{T}$- distributions for $\Xi_{cc}$, $\Xi_{bc}$ and $\Xi_{bb}$ production under the case of $p_T > 4$ GeV and $|y|< 1.5$ in Figs.(\ref{pt4yall},\ref{y15ptall}), where each curve includes the sum of all the mentioned S-wave diquark states. Fig.(\ref{y15ptall}) shows that production cross section of $\Xi_{bc}$ is smaller than that of $\Xi_{cc}$ in the lower $p_T$ region, however it will dominant over that of $\Xi_{cc}$ when $p_T \succsim 9$ GeV.

\section{Summary}

We have analyzed the hadronic production of $\Xi_{bc}$ via the dominant gluon-gluon fusion mechanism at LHC with the center-of-mass energy $\sqrt{S}=7$ TeV and $\sqrt{S}=14$ TeV respectively. For experimental usage, the total and the interested differential cross-sections have been estimated under typical cut conditions for the LHC detectors CMS, ATLAS and LHCb.

Numerical results show that about $1.7\times 10^7$ and $3.5\times10^9$ $\Xi_{bc}$ events per year can be produced for $\sqrt{S}=7$ TeV and $\sqrt{S}=14$ TeV under the condition of $p_T>4$ GeV and $|y|<1.5$. This indicates that $\Xi_{bc}$ can be observed and studied at LHC. Main uncertainties for the estimation have been discussed and a comparative study on the hadronic production of $\Xi_{cc}$, $\Xi_{bc}$ and $\Xi_{bb}$ at LHC with $\sqrt{S}=7$ TeV and $\sqrt{S}=14$ TeV have also been presented. As for the total production cross section under the case of $p_T<4$ GeV, we have $\sigma_{\Xi_{bc}}<\sigma_{\Xi_{cc}}$, however the differential cross-section of $\Xi_{bc}$ will dominant over that of $\Xi_{cc}$ when $p_T \succsim 9$ GeV.

In the above, we have not distinguished the light components in the baryon. More subtly, as for the production of $\Xi_{bc}$, after the formation of the heavy $(bc)$-diquark, it will grab a light anti-quark (with gluons when necessary) from the hadron collision environment to form a colorless double heavy baryon with the relative possibility for the light quark as $u :d :s \simeq 1:1:0.3$ \cite{pythia}, i.e. to form the baryons $\Xi_{bc}^{+}$, $\Xi_{bc}^{0}$ or $\Omega_{bc}^{0}$. More precisely, when the diquark $(bc)$ is produced, it will fragment into $\Xi_{bc}^{+}$ with $43\%$ probability, $\Xi_{bc}^{0}$ with $43\%$ probability and $\Omega_{bc}^{+}$ with $14\%$ probability accordingly. If enough $\Xi_{bc}$ events can be accumulated at LHC, then one may have chances to study the $\Xi_{bc}^{+,0}$ or $\Omega_{bc}^{0}$ separately from their decay products.

\hspace{1cm}

{\bf Acknowledgments:} This work was supported in part by the Fundamental Research Funds for the Central Universities under Grant No.CDJXS11102209, by Natural Science Foundation of China under Grant No.10805082 and No.11075225 and by Natural Science Foundation Project of CQ CSTC under Grant No.2008BB0298.

\end{document}